\documentclass[twocolumn,showpacs,amsmath,amssymb,pra,superscriptaddress]{revtex4-2}

\usepackage{graphicx}
\usepackage{dcolumn}
\usepackage{bm}
\usepackage{subfigure}
\usepackage{blindtext}

\begin{document}


\title {
Alternative treatment of relativistic effects in linear augmented plane wave \\
(LAPW) method: application to Ac, Th, ThO$_2$ and UO$_2$
}

\author {A. V. Nikolaev}

\affiliation{
Skobeltsyn Institute of Nuclear Physics, Moscow
State University, Vorob'evy Gory 1/2, 119234, Moscow, Russia
}

\author {U. N. Kurelchuk}

\affiliation{National Research Nuclear University MEPhI, Kashirskoe shosse 31, Moscow 115409, Russia}


\author {E. V. Tkalya}

\affiliation{P.N. Lebedev Physical Institute of the Russian Academy of Sciences, 119991, 53 Leninskiy pr., Moscow, Russia}

\affiliation{National Research Nuclear University MEPhI, Kashirskoe shosse 31, Moscow 115409, Russia}

\affiliation{Nuclear Safety Institute of RAS, Bol'shaya Tulskaya 52, Moscow 115191, Russia}

\affiliation{Institute of Nuclear and Radiation Physics, Russian Federal Nuclear Center-VNIIEF, 607188,
 Muzrukov Ave 10, Sarov, Nizhny Novgorod region, Russia}




\begin{abstract}
We examine the influence of the relativistic effects within the linear augmented plane wave method (LAPW) for solids and
propose a few alternative ways to accurately take them into account:
(1) we introduce new radial dependencies for LAPW (Bloch-type) basis functions, based on two actual radial solutions of
the Dirac equation for $j=l-1/2$ and $j=l+1/2$ states.
The proposed radial $6p$ functions receive more weight from the Dirac $p_{1/2}$ solution and, due to this, can on average correctly describe completely filled
$6p$ bands even without the additional $p_{1/2}$ local atomic function, as is done in the LAPW+$p_{1/2}$ method; (2)
the canonical LAPW matrix elements for the spherically symmetric component of
the potential, assuming non-relativistic radial wave functions, should be corrected;
(3) we argue that for a realistic spin-orbit (SO) energy splitting
of the semicore $6p-$states the spin-orbit interaction constant $\zeta(p)$ should be calculated with the $6p_{3/2}$ radial component,
because the value of $\zeta(p)$ obtained with the canonical mixing of the $6p_{1/2}$ and $6p_{3/2}$ components overestimates
the SO splitting.
Different ways of taking into account relativistic effects can change the equilibrium lattice constant up to 0.15 Å and the elastic modulus up to 26 GPa.
We find that in the full treatment of the spin-orbit coupling
UO$_2$ has a small gap of forbidden states ($0.2-0.4$~eV) at the Fermi level, which persists for all $\vec{k}-$vectors and, therefore, UO$_2$
should be classified as a semimetal. We also discuss the peculiarities of the electron band structure of actinium, which result in
an overestimation of its lattice constant.
\end{abstract}


\maketitle

\section{Introduction}
\label{sec:int}

Nowadays band structure calculations become a powerful tool of investigation of complex materials proving its efficiency for many solids
and capable of predicting their properties.
The accuracy of such calculations increases with every year and there is a constant demand for even better precision and performance.
As shown in benchmark calculations \cite{bench},  various band structure methods generally result in the same or close final results (for example, the equilibrium lattice constants, bulk moduli etc.).
Neverthelss, there is a class of materials where the description of solids is less certain and encounters difficulties.
Such materials include, in particular, heavy elements -- actinides, situated in the end of the Mendeleev Periodic Table, which heave eighty or more core electrons,
experiencing relativistic effects.
In the literature there are several successful studies of band structure of actinides.
First works were
based on the linear augmented muffin-tin method (LMTO) \cite{review-LMTO,Pet,Th-LMTO,ThO2-LMTO},
followed by studies \cite{Jones,Pen}
carried out with the full potential linear augmented plane wave method (FLAPW) \cite{KA,wien2k,blapw,Nik2},
considered as one of the most precise band structure methods.
Additional complexity of the subject is related to the fact that the electrons, belonging to the incompletely filled $5f$ shell of actinides
exhibit competition between itineracy and localization \cite{Lander,Skriv}.
This competition can lead to nontrivial magnetic and other correlation effects \cite{Sarrao,Heat,Petit1,Korz,Sod1}.

In the canonical FLAPW approach one uses so called scalar relativistic approach
based on the works of Koelling and Harmon in \cite{KH}, and MacDonald, Pickett and Koelling in Ref.\cite{MPK}.
(We consider it in detail below in Sec. \ref{sub:basis}.)
In a further development, it was proposed that the FLAPW basis set be enriched with local atomic functions \cite{Kunes,Vona},
which can be fully relativistic, i.e. taken from the solution to the radial Dirac equation.
In Ref.\ \cite{Kunes} the relativistic $p_{1/2}$ local orbitals were added in the second variation step of
the FLAPW calculation of elemental thorium (FLAPW$+p_{1/2}$),
which turned out to significantly improve the stability and precision of band calculations.
Two variational procedures are often used as a time-saving computational scheme in the FLAPW method
when the spin-orbit (SO) coupling is included.
In the first step only the scalar relativistic part of the Hamiltonian is diagonalized,
whereas in the second variation step the SO coupling matrix is constructed and then diagonalized
in a smaller basis set, consisting of a limited number of low lying eigenfunctions obtained on the first step
and additional local atomic orbitals \cite{Kunes,Vona}. In Ref.\ \cite{Kunes} only the $p_{1/2}$ relativistic
atomic functions were used, while in Ref.\ \cite{Vona} the method was extended to include other local
relativistic functions and their combinations.

The proposed corrections \cite{Kunes,Vona} through the relativistic local orbitals in the second variation step
being effective in practice, are based on the idea to increase the convergence and effectiveness of
the basis set. It does not improve the relativistic characteristics of canonical LAPW (i.e. Bloch-like) band functions.
As shown later in Sec.\ \ref{sec:meth} the present scalar relativistic LAPW
method \cite{KH,MPK}, when applied to heavy elements like actinides and transactinides,
demonstrates deviations of the averaged relativistic radial functions from the canonical
scalar relativistic ones (Sec.\ \ref{sub:basis})
and require small corrections in expressions for matrix elements (Sec.\ \ref{sub:mel}).
In addition, the second variation step performed on a small number of secondary basis functions
(which is usually considerably smaller the full basis set) can be considered as a perturbative treatment of the spin-orbit coupling \cite{Jones}.

The aim of the present study is to increase the accuracy of the relativistic effects within the full electron full potential
LAPW method (FLAPW) method for heavy elements as much as possible,
while keeping the general scheme of the method unchanged.
For that purpose (1) we introduce new basis functions, obtained from two independent
solutions of the Dirac equation, which differ from the canonical LAPW basis functions.
In modern LAPW method one deals with two types of basis functions: Bloch-like LAPW basis functions and
the atomic-like local orbitals (LO), which can be chosen to be fully relativistic (as e.g. the $p_{1/2}$ mentioned earlier \cite{Kunes,Vona}).
{\it In our work the new type of radial basis functions concerns the Bloch-type LAPW basis functions.}
For the very important $6p$ semicore states, the proposed new radial basis functions gain more weight of the $6p_{1/2}$ Dirac component
in comparison with the canonical functions, and, as a result of that, they can describe the $6p$ bands even without addition the $p_{1/2}$ local
atomic function. Therefore, {\it the present approach can be considered as an alternative to the standard (FLAPW$+p_{1/2}$) scheme}.
We discuss the peculiarity of this treatment in Sec.\ \ref{sub:basis}.
We do not consider here the choice of local orbitals in our method, which can also be added to improve the quality of the basis set,
since it is a separate problem \cite{Nik2,Michal,Karsai}.

Further, (2) we reconsider the matrix elements of the method,
explicitly avoiding the use of hidden non-relativistic relations;
(3) we correct the calculation of the spin-orbit (SO) coupling constant for $6p$ semicore states,
based on the comparison the energy splittings between $6p_{1/2}$ and $6p_{3/2}$ components.
In the present study we do not apply the second variation procedure for the SO-coupling.
We use the direct treatment of the SO coupling in the full LAPW basis set
thereby avoiding the approximations associated with the second variation step \cite{blapw}.
These effects are considered and discussed in Sec.\ \ref{sub:basis}, Sec.\ \ref{sub:mel} and Sec.\ \ref{sub:soc-p}, correspondingly.
In Sec.\ \ref{sec:appl} we briefly review the results of our calculations for Ac, Th, ThO$_2$
and UO$_2$, and finally in Sec.\ \ref{sec:con} we discuss main conclusions and findings of our work.
In our study we use various variants of the DFT functionals (see Sec.\ \ref{sec:appl} below),
which allows us to test the accuracy of the calculations.

\section{Method}
\label{sec:meth}

In the LAPW method \cite{KA,wien2k,blapw}, widely used for studies of bulk materials, the space is partitioned in the region inside the nonoverlapping muffin-tin (MT) spheres and
the interstitial region (IR).
The basis functions $\phi_j(\vec{k},\,\vec{R})$, where $j=1,2,...,N_b$, are given by
\begin{equation}
  \phi_j(\vec{k},\,\vec{R}) = \left\{ \begin{array}{ll} v^{-1/2} \, exp(i(\vec{k} + \vec{K}_j) \vec{R}), & \vec{R} \in IR   \\
                                                    \sum_{l,m} {\cal R}_{l,m}^{j,\alpha}(r,\, E_l)\, Y_{l,m}(\hat{r}), & \vec{R} \in MT(\alpha)  \end{array} \right.
  \label{m1}
\end{equation}
where $\vec{K}_j$ refers to the reciprocal lattice vector $j$, $v$ is the unit cell volume,
$Y_{l,m}$ are spherical harmonics \cite{BC} and the radial part is given by
\begin{equation}
  {\cal R}_{l,m}^{j,\alpha}(r,\, E_l) = A^{j,\alpha}_{l,m}\, u_l(r, E_l) + B^{j,\alpha}_{l,m}\, \dot{u}_l(r, E_l) .  \quad
 \label{m2}
\end{equation}
Here the index $\alpha$ refers to the type of atom (or
MT-sphere) in the unit cell, the radius $r$ is counted from the
center $\vec{R}_{\alpha}$ of the sphere $\alpha$ (i.e.
$\vec{r}=\vec{R}-\vec{R}_{\alpha}$). Radial functions $u_{l}(r,E_l)$ are solutions
in the spherically averaged crystal
potential computed at the linearization energy $E_l$, and
$\dot{u}_l(r,E_l)$ is the derivative of $u_{l}$ with respect to
$E$ at $E_l$. The coefficients $A^{j,\alpha}_{l,m}$ and $B^{j,\alpha}_{l,m}$ are
found from the condition that the basis function $\phi_j$ is
continuous with continuous derivative at the sphere boundary, i.e. at
$r=R_{MT}^{\alpha}$ ($R_{MT}^{\alpha}$ is the radius of the
MT-sphere $\alpha$).
The coefficients $A^{j,\alpha}_{l,m}$ and $B^{j,\alpha}_{l,m}$ in Eq.\ (\ref{m2}) are related to the standard LAPW quantities $a^{j}_{l}$, $b^{j}_{l}$,
expressed only through the spherical Bessel functions $j_l$ and the radial solution $u_l$ (and its derivatives) at $r=R_{MT}^{\alpha}$.

\subsection{Explicitly averaged radial basis wave functions}
\label{sub:basis}

Initially, the functions $u_{l}(r,E_l)$ in Eq.\ (\ref{m2}) were considered as the solutions of the Schr\"{o}dinger equation
in the spherically symmetric ($L=0$) component of the total potential.
Later, it appeared that some relativistic effects can be included in the so called scalar relativistic approach \cite{KH,MPK}.
Below we discuss the canonical radial functions introduced by Koelling and Harmon in \cite{KH}, later justified by the procedure
described by MacDonald, Pickett and Koelling in Ref.\cite{MPK}, and compare them with new radial functions that are more closely
related to the Dirac solutions.

The standard LAPW radial basis functions are defined by an average
\begin{eqnarray}
   P^{av}_l(r) = \frac{l}{2 l + 1} P_l(r) + \frac{l + 1}{2 l + 1} P_{-l-1} (r),
\label{i1a}
\end{eqnarray}
where $r$ is radius and $P_l$, $P_{-l-1}$ are the large ($L$) components of the Dirac solutions $P_{\kappa^L}$ for $\kappa^L = l$
($j = l - 1/2$), and $\kappa^L = -l-1$ ($j = l + 1/2$),
correspondingly. (Here $\kappa^L$ stands for the index $\kappa$ of 2-spinors for the large component \cite{Rel-book}.)
However, in practice the large components $P_l$ and $P_{-l-1}$ are not calculated.
In Ref.\ \cite{MPK} assuming that
\begin{eqnarray}
   \frac{d}{d r} (\delta P(r)) = \frac{d}{d r} \left( P_{-l-1} (r) -  P_l(r) \right) = 0 ,
\label{i1b}
\end{eqnarray}
an effective system for two coupled differential equations was derived.
Then the LAPW radial basis function is
\begin{eqnarray}
   P^{KH}_l(r) = \left. P^{av}_l(r) \right|_{\delta P'(r) = 0} ,
\label{i1c}
\end{eqnarray}
i.e. the function (\ref{i1a}), provided that the condition (\ref{i1b}) is fulfilled.
The second auxiliary function is an averaged small component $Q^{KH}_l(r)$ \cite{MPK}, given by Eq.~(3b) of \cite{MPK},
but except in the system of differential equations, it is not used.
As a result, $P^{KH}_l$, $Q^{KH}_l$ depend only on $l$ and include some relativistic effects.
Although this approach has proved being efficient and practical, it has serious drawbacks when applied to heavy elements.
In particular, Eq.~(\ref{i1b}) is {\it an uncontrolled approximation}, and the averaging for $Q^{KH}_l$ in Eq.~(\ref{i1c}) is a formal procedure,
for angular two-spinors $\xi_{-\kappa, m}$, associated with
the Dirac small components $Q_{-l}$ and $Q_{l+1}$, have different angular dependencies \cite{Rel-book}.

%
\begin{figure}
\resizebox{0.45\textwidth}{!} {
\includegraphics{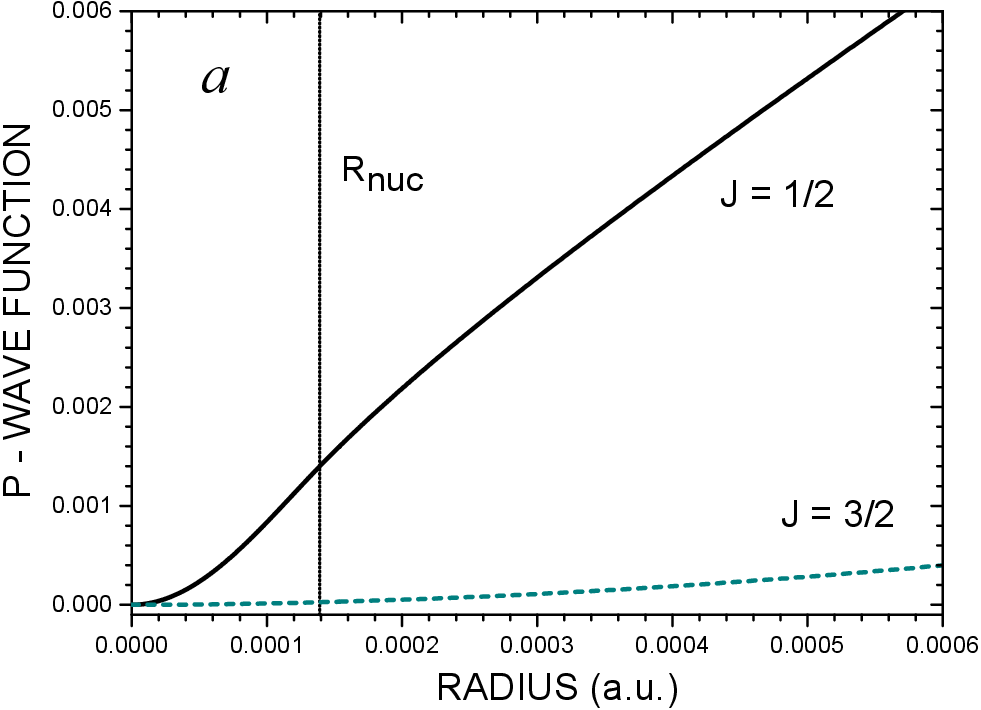}}
\resizebox{0.45\textwidth}{!} {
\includegraphics{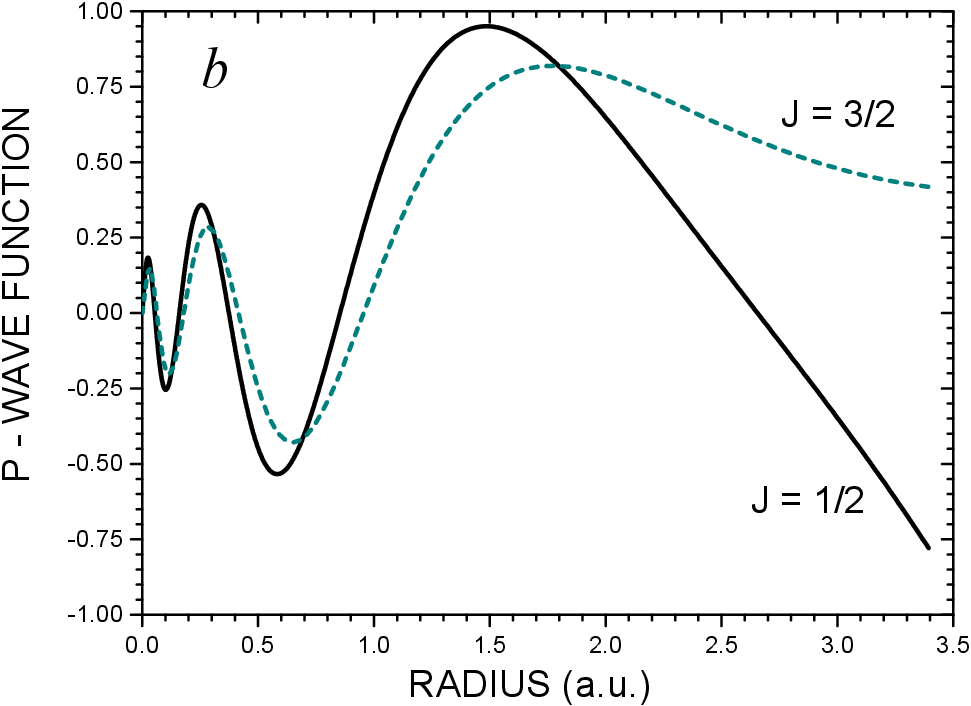}
}

\caption{
Radial Dirac functions $P_{j=1/2}(r)$ and $P_{j=3/2}(r)$ of the $6p_{1/2}$ and $6p_{3/2}$ semicore states of thorium,
(a) close to the nuclear region, and
(b) inside the MT-sphere.
} \label{fig1}
\end{figure}
%

\begin{figure}
\resizebox{0.45\textwidth}{!} {
\includegraphics{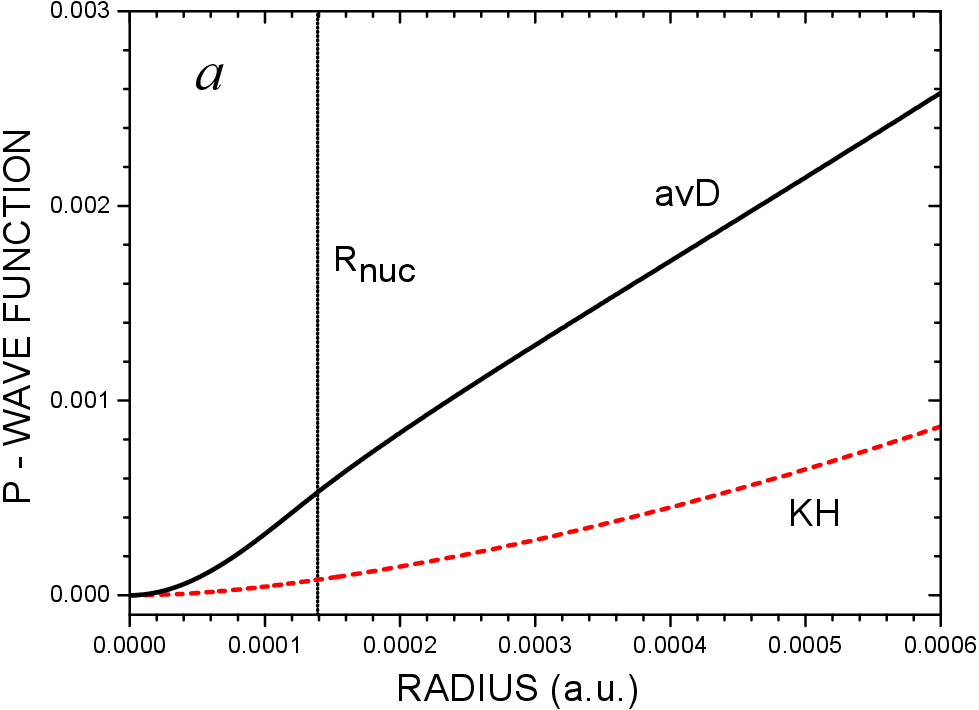}}
\resizebox{0.45\textwidth}{!} {
\includegraphics{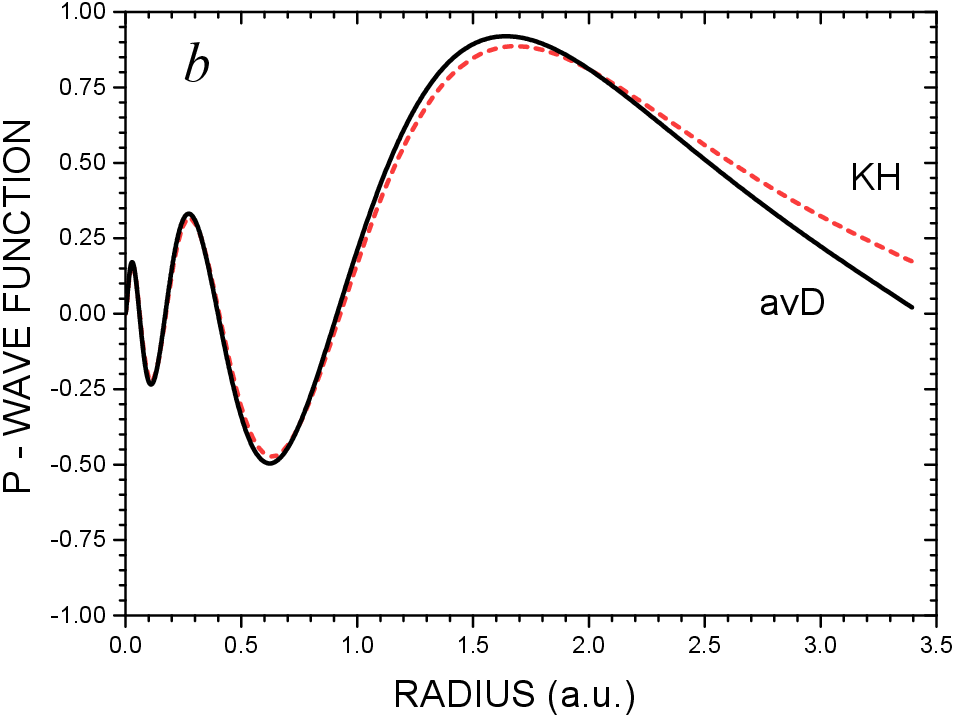}
}

\caption{
Radial basis function $P^{av}_{\ell=1}(r)$ (avD) and the canonical KH radial basis function $P^{KH}_{l=1}(r)$ \cite{KH,MPK}
of the semicore $6p$ states of thorium,
(a) close to the nuclear region, and
(b) inside the MT-sphere.
} \label{fig2}
\end{figure}

To make the discussion on the radial part more concrete, we start with the case of the face centered cubic (fcc) lattice of elemental thorium
(with the PBE exchange correlation functional \cite{PBE}) and consider the $6p_{1/2}$ and $6p_{3/2}$ valence (semicore) states.
In Fig.\ \ref{fig1} we plot the radial dependencies $P_{j=1/2}(r)$ and $P_{j=3/2}(r)$ for the large components of $6p_{1/2}$ and $6p_{3/2}$ states
at the same energy obtained by solving the Dirac equation in the self-consistent spherical potential.
One clearly sees the different shape of large components both at the
large radii close to the MT-radius and at the neighborhood of the nuclear region. Then we calculate the average radial function $P^{av}_{\ell=1}(r)$ for the $6p$-states
using Eq.\ (\ref{i1a}) explicitly. This numerically averaged function $P^{av}_{\ell=1}(r)$, as well as
the conventional LAPW function $P_{l=1}^{KH}$, obtained by solving the KH differential equations \cite{KH,MPK}, are reproduced in Fig.\ \ref{fig2}.
Note that the numerically averaged radial function $P^{av}_{\ell=1}$ remains different from $P_{l=1}^{KH}$ both at small and large radii,
reflecting the approximate character of Eq.\ (\ref{i1b}) and Eq.\ (\ref{i1c}).
In the following instead of $P_{l=1}^{KH}(r)$, and the corresponding energy derivative radial function $\dot{P}^{KH}_{\ell=1}(r) = \partial P^{KH}_{\ell=1}(r) / \partial E$, required by the LAPW method, we suggest to use the explicitly averaged functions $P^{av}_{\ell=1}$, $\dot{P}^{av}_{\ell=1}$ as the $6p$ LAPW basis set.
That is, we calculate first the Dirac functions $P_l$ and $P_{-l-1}$ ($l = 1$), after which we obtain the radial basis function $P^{av}_l$
according to Eq.~(\ref{i1a}). A very important advantage of this scheme is that it avoids the use of the uncontrolled assumption (\ref{i1b}).
%
%

As shown in Fig.\ \ref{fig2}a in the vicinity of the Th nucleus $P^{av}_{\ell=1}$ considerably exceeds $P_{l=1}^{KH}$, indicating that $P^{av}_{\ell=1}$ gains more weight of
the $P_{1/2}$ Dirac component, Fig.\ \ref{fig1}a, than $P_{l=1}^{KH}$. The same conclusion can be drawn from the behavior of $P^{av}_{\ell=1}$ at
larger radii ($r \sim 3-3.5$~a.u.) where again $P^{av}_{\ell=1}$ lies closer to the $P_{1/2}$ radial solution shown in Fig.\ \ref{fig1}b.
As a result, new averaged $6p$ radial function $P^{av}_{\ell=1}$, paired with the complimentary radial function $\dot{P}^{av}_{\ell=1}$,
reproduces on average the sum of two components ($p_{1/2}$ and $p_{3/2}$) much better than $P_{l=1}^{KH}$ with $\dot{P}_{l=1}^{KH}$ and can describe the electron density
of the $6p$ bands correctly. This is demonstrated by our calculations in Sec.\ \ref{sec:appl} with the new basis functions $P^{av}_{\ell=1}$ and $\dot{P}^{av}_{\ell=1}$.
Note, that these calculations have been performed without the additional $p_{1/2}$ local atomic function whereas
in the canonical basis set (KH $6p$ radial functions $P_{l=1}^{KH}$, which is closer to $P_{3/2}$) the weight of the $p_{1/2}$ states is so small that the use of the $p_{1/2}$ local function is mandatory.

Similarly to the explicit procedure of averaging $p_{1/2}$ and $p_{3/2}$, described above, we can introduce new radial functions for $d-$ and $f-$ (and high $\ell$) states
by using Eq.\ (\ref{i1a}) for the independently calculated $j = l - 1/2$ and $j = l + 1/2$ Dirac radial functions.
In the following we will refer to these directly averaged radial functions $P^{av}_{\ell}$ (large components) of the Dirac solutions as
new basis functions (denoted as the basis set avD)
comparing their performance with the standard KH-basis functions $P^{KH}_{l}$ (the KH basis set).
The most important difference with the $6p$ states is that for $d-$ and $f-$ states two Dirac radial functions ($j = l - 1/2$ and $j = l + 1/2$)
lie close to each other.
To characterize quantitatively the difference between $P^{av}_{\ell}$ and $P_{l}^{KH}$ ($\dot{P}^{av}_{\ell}$ and $\dot{P}_{l}^{KH}$) for $\ell > 0$
we introduce the deviation quantities $\triangle P_l$ and $\triangle \dot{P}_l$, defined as
\begin{subequations}
\begin{eqnarray}
     \triangle P_l &=& \sqrt{\int_0^{R_{MT}} (P^{av}_{\ell}(r) - P_{l}^{KH}(r))^2} , \label{i3a} \\
     \triangle \dot{P}_l &=& \sqrt{\int_0^{R_{MT}} (\dot{P}^{av}_{\ell}(r) - \dot{P}_{l}^{KH}(r))^2} , \label{i3b}
\end{eqnarray}
\end{subequations}
Calculated values of $\triangle P_l$ and $\triangle \dot{P}_l$ for various elements and compounds are listed in Table \ref{tab1}.
Note that the large component of $s_{1/2}$ functions in both treatments coincides,
i.e. $P^{av}_{\ell=0} = P_{l=0}^{KH}$, $\dot{P}^{av}_{\ell=0} = \dot{P}_{l=0}^{KH}$.

Inspection of Table \ref{tab1} shows that the largest difference between two functions (more than 10{\%} of its norm) is found for $6p$-basis states of actinides.
The differences between the $d-$ and $f-$ functions are of the order of only $10^{-3}$, but one should have in mind that Eq.\ (\ref{i3a}) and Eq.\ (\ref{i3b})
are integral. As we will see in Sec.\ \ref{sub:soc-p} even this small difference matters for the calculation of SO coupling constants $\zeta$,
because in atomic units
\begin{eqnarray}
     \zeta =  \frac{1}{2c^2} \left\langle \frac{1}{r} \frac{dV}{dr} \right\rangle_{av} \sim \left\langle \frac{Z}{r^3} \right\rangle_{av} , \label{i4}
\end{eqnarray}
where $V(r)$ is the Coulomb potential.
Therefore in this case the neighborhood of the nucleus, where the differences are visible, Fig.\ \ref{fig3}, contributes
with considerably larger weight than the other regions.
In general, however, the explicitly averaged radial functions (avD) for $6d$ ($P^{av}_{\ell=2}$), $5f$ ($P^{av}_{\ell=3}$) and higher $\ell-$states demonstrate
a much more close correspondence with the KH-radial functions $P_{l=2}^{KH}$, $P_{l=3}^{KH}$, etc., because of weak presence of these functions in the nuclear region.
We will return to this problem in Sec.\ \ref{sub:soc-p} below.
Differences between the avD and KH radial basis functions for light elements such as oxygen in ThO$_2$ or UO$_2$ are negligible, Table \ref{tab1}.

%
\begin{table}
\caption{
Deviations $\triangle P_l$ and $\triangle \dot{P}_l$, Eq.\ (\ref{i3a}) and Eq.\ (\ref{i3b}),
between two radial basis functions: the canonical KH $P_{l}^{KH}$ ($\dot{P}_{l}^{KH}$) and the explicitly averaged Dirac (avD)
functions $P_{\ell}^{av}$ ($\dot{P}_{\ell}^{av}$), Eq.\ (\ref{i1a}), in fcc Th, fcc Ac, bcc Np, cubic ThO$_2$ [Th(2), O(2)]
and UO$_2$ [O(3)] for $l > 0$; $s-$functions coincide ($\triangle P_{l=0} < 10^{-10}$).
\label{tab1} }
\begin{ruledtabular}
\begin{tabular}{c  l  c  c  c  c}
     &       & $p$ & $d$ & $f$ &  $\ell > 3$ \\
\tableline
  Th   & $\triangle P_l$       & 0.108 & 1.1$\cdot 10^{-3}$ & 1.5$\cdot 10^{-3}$ & $<$2.3$\cdot 10^{-8}$ \\
  Th   & $\triangle \dot{P}_l$ & 0.148 & 1.4$\cdot 10^{-3}$ & 3.9$\cdot 10^{-3}$ & $<$1.7$\cdot 10^{-8}$ \\
  Th(2)& $\triangle P_l$       & 0.033  & 4.3$\cdot 10^{-4}$ & 3.4$\cdot 10^{-4}$ & $<$1.0$\cdot 10^{-7}$ \\
  Th(2)& $\triangle \dot{P}_l$ & 0.012  & 3.2$\cdot 10^{-4}$ & 2.8$\cdot 10^{-4}$ & $<$4.2$\cdot 10^{-8}$ \\
  O(2) & $\triangle P_l$       & 2.9$\cdot 10^{-6}$  & 2.2$\cdot 10^{-9}$  & 5.5$\cdot 10^{-10}$ & $<$3.3$\cdot 10^{-10}$ \\
  O(2) & $\triangle \dot{P}_l$ & 2.8$\cdot 10^{-6}$  & 9.4$\cdot 10^{-10}$ & 2.3$\cdot 10^{-10}$ & $<$2.9$\cdot 10^{-10}$ \\
  U    & $\triangle P_l$       & 0.040 & 5.6$\cdot 10^{-4}$ & 6.3$\cdot 10^{-4}$ & $<$1.5$\cdot 10^{-7}$ \\
  U    & $\triangle \dot{P}_l$ & 0.014 & 3.3$\cdot 10^{-4}$ & 4.7$\cdot 10^{-4}$ & $<$5.9$\cdot 10^{-8}$ \\
  O(3) & $\triangle P_l$       & 2.9$\cdot 10^{-6}$  & 2.3$\cdot 10^{-9}$  & 5.5$\cdot 10^{-10}$ & $<$3.3$\cdot 10^{-10}$ \\
  O(3) & $\triangle \dot{P}_l$ & 2.6$\cdot 10^{-6}$  & 9.7$\cdot 10^{-10}$ & 2.1$\cdot 10^{-10}$ & $<$2.9$\cdot 10^{-10}$ \\
  Ac   & $\triangle P_l$       & 0.094   & 9.2$\cdot 10^{-4}$ & 5.3$\cdot 10^{-4}$ & $<$1.5$\cdot 10^{-8}$ \\
  Ac   & $\triangle \dot{P}_l$ & 0.170   & 1.2$\cdot 10^{-3}$ & 1.5$\cdot 10^{-3}$ & $<$1.2$\cdot 10^{-8}$ \\
  Np   & $\triangle P_l$       & 0.081   & 9.3$\cdot 10^{-4}$ & 1.8$\cdot 10^{-3}$ & $<$8.1$\cdot 10^{-8}$ \\
  Np   & $\triangle \dot{P}_l$ & 0.038   & 6.4$\cdot 10^{-4}$ & 1.9$\cdot 10^{-3}$ & $<$4.3$\cdot 10^{-8}$ \\
\end{tabular}
\end{ruledtabular}
\end{table}

\subsection{Correction of LAPW matrix elements}
\label{sub:mel}

The use of relativistic basis functions requires a modification of some matrix elements
which are valid only in the non-relativistic limit.
In particular, the matrix elements of the $L=0$ component of the potental,
as written in Eq.\ (16a) and Eq.\ (16b) of Ref.\ \cite{KA} are exact only in the nonrelativistic limit.
The existence of this limitation is due to the fact that in deriving the expressions for the matrix elements, the equality
\begin{eqnarray}
 R_{MT}^2 ( \dot{u}_l(R_{MT}) \, u'_l(R_{MT}) - \dot{u}'_l(R_{MT})\, u_l(R_{MT}) ) = 1, \quad
 \label{cor1}
\end{eqnarray}
is used. (Here, as before, the energy derivative $\dot{u}(r) = \partial u(r)/\partial E$ is defined in Rydberg energy units.)
Eq.\ (\ref{cor1}), explicitly quoted in Ref.\ \cite{KA} as Eq.\ (4), is exact only for the radial component $u_l$ of
the Schr{\"o}dinger equation in the spherically symmetric potential.

In general, the expression on the left hand side of Eq.\ (\ref{cor1}) deviates from unity for the effective radial components $P^{av}_{\ell}$, $\dot{P}^{av}_{\ell}$ and
$P^{KH}_l$, $\dot{P}^{KH}_l$, described in Sec.\ \ref{sub:basis}, because they are obtained from the Dirac equation.
To illustrate this, in Table~\ref{tab2} we reproduce the values of the deviation factor
\begin{eqnarray}
  F(l) = R_{MT}^2 ( \dot{P}_l(R_{MT}) \, P'_l(R_{MT}) \nonumber \\
  - \dot{P}'_l(R_{MT})\, P_l(R_{MT}) ) - 1 , \label{cor2}
\end{eqnarray}
for the avD and KH radial basis functions. For the nonrelativistic (Schr{\"o}dinger) functions
we have $F(l) \equiv 0$. In practice, as shown in Table \ref{tab2} we find $F(l) \neq 0$.
The deviations are the largest ($F(p)=-0.13$) for the 6$p$ radial functions in the avD-basis set.
In the KH basis set $F(l)$ are smaller, reaching only the value $F(p) \approx 7 \cdot 10^{-4}$ for $6p-$states of Ac.
However, even such deviations can lead to a sizeable inaccuracy in determination of the equilibrium lattice constants
and bulk moduli, and should be avoided.

The inequality $F(l) \neq 0$ requires amendments to some expressions of the LAPW method.
The corrections involve the precise determination of the $a_l$ and $b_l$ coefficients and the matrix elements for
the spherically symmetric component of the potential.
In particular, Eq.\ (10b) and Eq.\ (10d) of Ref.\ \onlinecite{KA} should be replaced with
\begin{subequations}
\begin{eqnarray}
  a^n_l =  \frac{1}{\triangle} ( j'_l(k_n R_{MT}) \, \dot{P}_l - j_l(k_n R_{MT}) \, \dot{P}'_l ) , \label{cor3a} \\
  b^n_l = \frac{1}{\triangle} ( j_l(k_n R_{MT}) \, P'_l - j'_l(k_n R_{MT}) \, P_l ) , \label{cor3b}
\end{eqnarray}
\end{subequations}
where
\begin{eqnarray}
  \triangle =  R_{MT}^2 (\dot{P}_l \, P'_l - P_l \, \dot{P}'_l) \neq 1 . \label{cor4}
\end{eqnarray}

The value for $\gamma^l$ given in Eq.\ (16b) of Ref.\ \cite{KA}, should also be rewritten.
In the notation of Ref.\ \cite{KA} the following symmetric form can be obtained
\begin{eqnarray}
  \gamma^l &=& \frac{1}{2} \left\{ a_l(\vec{k}_n) b_l(\vec{k}_m) + a_l(\vec{k}_m) b_l(\vec{k}_n) \right. \nonumber \\
  & & + \left. \frac{1}{R^2_{MT}} (j'_l(n) j_l(m) + j_l(n) j'_l(m)) \right\} .
  \label{cor5}
\end{eqnarray}
Here the second part with the Bessel functions comes from the MT-sphere boundary integration of the kinetic energy performed for
the symmetrization of the expression for the matrix elements of kinetic energy, i.e. it appears due to the
replacement of $\vec{k}_n \vec{k}_n U$ (or $\vec{k}_m \vec{k}_m U$) with $\vec{k}_n \vec{k}_m U$, Eq.\ (16a) of \cite{KA}.
%
\begin{table}
\caption{
Deviation of the factors $F(l)$ from zero, Eq.\ (\ref{cor2}), for the avD, Eq.\ (\ref{i1a}),
and the canonical KH \cite{KH,MPK} basis functions in fcc Th, fcc Ac, bcc Np, cubic ThO$_2$ [Th(2), O(2)]
and UO$_2$ [O(3)], underlying the importance of the corrections
for LAPW matrix elements, Eqs.\ (\ref{cor3a}), (\ref{cor3b}) and Eq.\ (\ref{cor5}).
\label{tab2} }
\begin{ruledtabular}
\begin{tabular}{l  l  c  c  c  c}
     & basis &  $s$  & $p$ & $d$ & $f$  \\
\tableline
  Th    & avD & $1.2 \cdot 10^{-4}$ & $-0.129$            & $4.0 \cdot 10^{-4}$ & $1.8 \cdot 10^{-4}$ \\
  Th    & KH  & $1.1 \cdot 10^{-4}$ & $2.0 \cdot 10^{-4}$ & $7.9 \cdot 10^{-5}$ & $1.4 \cdot 10^{-4}$ \\
  Th(2) & avD & $5.1 \cdot 10^{-4}$ & $3.3 \cdot 10^{-3}$ & $8.5 \cdot 10^{-4}$ & $2.4 \cdot 10^{-4}$ \\
  Th(2) & KH  & $5.1 \cdot 10^{-4}$ & $2.4 \cdot 10^{-4}$ & $2.1 \cdot 10^{-4}$ & $2.1 \cdot 10^{-4}$ \\
  O(2)  & avD & $6.7 \cdot 10^{-5}$ & $7.3 \cdot 10^{-5}$ & $1.1 \cdot 10^{-5}$ & $<1e^{-6}$ \\
  O(2)  & KH  & $6.6 \cdot 10^{-5}$ & $7.1 \cdot 10^{-5}$ & $1.4 \cdot 10^{-5}$ & $1.3 \cdot 10^{-5}$ \\
  U     & avD & $5.2 \cdot 10^{-4}$ & $1.5 \cdot 10^{-3}$ & $8.3 \cdot 10^{-4}$ & $2.6 \cdot 10^{-4}$ \\
  U     & KH  & $5.2 \cdot 10^{-4}$ & $2.6 \cdot 10^{-4}$ & $2.1 \cdot 10^{-4}$ & $2.9 \cdot 10^{-4}$ \\
  O(3)  & avD & $6.2 \cdot 10^{-5}$ & $7.8 \cdot 10^{-5}$ & $1.3 \cdot 10^{-5}$ & $2.0 \cdot 10^{-6}$ \\
  O(3)  & KH  & $6.5 \cdot 10^{-5}$ & $7.7 \cdot 10^{-5}$ & $1.6 \cdot 10^{-5}$ & $1.4 \cdot 10^{-5}$ \\
  Ac    & avD & $1.0 \cdot 10^{-4}$ & $-0.126$            & $3.6 \cdot 10^{-4}$ & $1.6 \cdot 10^{-4}$ \\
  Ac    & KH  & $7.5 \cdot 10^{-4}$ & $6.8 \cdot 10^{-4}$ & $4.5 \cdot 10^{-4}$ & $3.9 \cdot 10^{-4}$ \\
  Np    & avD & $2.6 \cdot 10^{-4}$ & $-0.039$            & $5.5 \cdot 10^{-4}$ & $-2.9 \cdot 10^{-4}$ \\
  Np    & KH  & $2.6 \cdot 10^{-4}$ & $2.4 \cdot 10^{-4}$ & $1.3 \cdot 10^{-4}$ & $3.1 \cdot 10^{-4}$ \\
\end{tabular}
\end{ruledtabular}
\end{table}

\subsection{Calculated spin-orbit coupling constants, special treatment for $6p-$states}
\label{sub:soc-p}

In this section we consider how the new (avD) basis functions affect the values of the spin-orbit (SO) energy splittings.
As a test exercise we first calculate spin-orbit coupling (SOC) constants and energy splittings for relativistic atoms of Ac, Th, U and Np
with the PBE variant of DFT for exchange and correlations \cite{PBE}.
We have performed two different sets of atomic calculations. In the first case, for the core shells we used the Dirac equations while
for all valence states the KH \cite{KH} and MPK differential equations \cite{MPK}
were employed. These are the same differential equations for $P_l^{KH}$ functions, which are used in the LAPW method.
In the second case, the Dirac equations were used for both core and valence states, which is an exact treatment. The energy splittings between the
$j = l - 1/2$ and $j = l + 1/2$ valence ($7d$-, $5f$- and $6p$-) states ($\triangle E$ in Tables \ref{tab4}--\ref{tab3}) then serve as reference values.

In atomic units the SO coupling constant $\zeta(l)$, defined by the radial function $P_l(r)$, can be found as
\begin{eqnarray}
   \zeta(l) = \frac{1}{2 c^2}  \int_0^{\infty} dr\, P_l^2(r)\, \frac{1}{r} \frac{dV}{dr} ,  \label{so1}
\end{eqnarray}
where $V(r)$ is the radial dependence of the Coulomb potential.
The corresponding SO operator is
\begin{eqnarray}
 H^{SO} = \zeta(l)\, \hat{L} \hat{S}  ,  \label{so1B}
\end{eqnarray}
with energy splitting
\begin{eqnarray}
 \triangle_{SO}(l) = \zeta(l) \frac{2 l + 1}{2}  .  \label{so1C}
\end{eqnarray}
As $P_l$ we consider either $P_{\ell}^{av}$ (the avD basis) or $P_{l}^{KH}$ (the KH basis).
Comparing $\triangle_{SO}(l)$, Eq.\ (\ref{so1C}), with the actual energy splitting $\triangle E(l)$, obtained by solving the Dirac atomic eigenproblem directly,
we can conclude which basis set (avD or KH) gives a better description of the SO splitting.
For the KH-basis, during the self-consistent procedure all core electron shells were obtained according to the fully relativistic Dirac approach
whereas all valence electron shells according to the KH-equations \cite{KH,MPK}, as is done in LAPW calculations.
The calculated values of the SO coupling for $6d$ states are listed in Table \ref{tab4},
for $5f$ states in Table \ref{tab5} and for $6p$ states in Table \ref{tab3}.
Since for the avD-basis set we calculate individual Dirac radial components, we also quote the individual SO couplings $\zeta(l)$ for them, i.e.
for $d_{3/2}$, $d_{5/2}$ states in Table \ref{tab4}, for $f_{5/2}$, $f_{7/2}$ in Table \ref{tab5} and for $p_{1/2}$, $p_{3/2}$ states in Table~\ref{tab3}.

\begin{figure}
\resizebox{0.45\textwidth}{!} {
\includegraphics{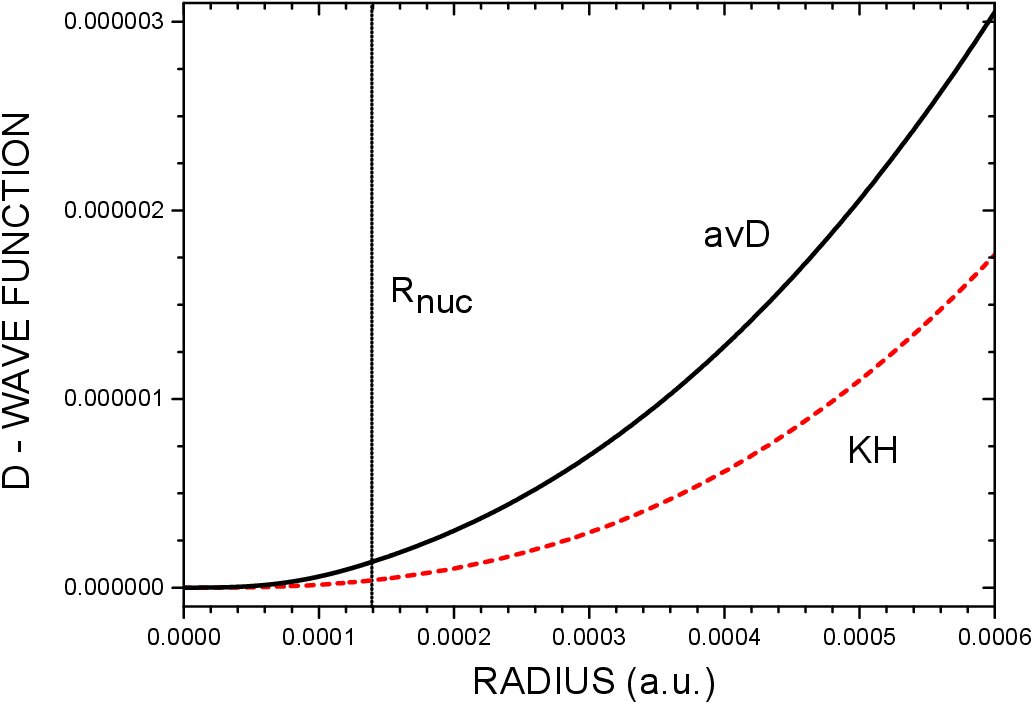}}
\resizebox{0.45\textwidth}{!} {
\includegraphics{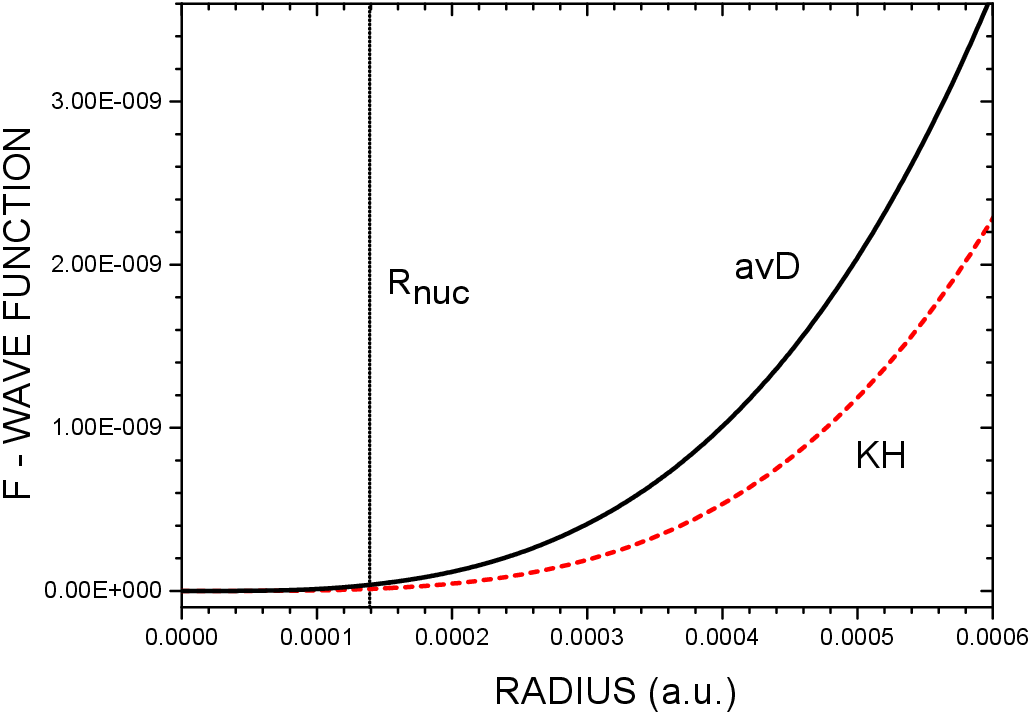}
}

\caption{
Explicitly averaged radial functions $P^{av}_{\ell}(r)$ (a) for $6d$ ($\ell=2$), and (b) for $5f$ ($\ell=3$) states and the
corresponding canonical KH radial function $P^{KH}_{l}(r)$ \cite{KH,MPK}
close to the nuclear region.
} \label{fig3}
\end{figure}

Comparing $\triangle_{SO}(l)$ with $\triangle E(l)$ in Table \ref{tab4} for $d-$states and Table \ref{tab5} for $f-$states shows that in all cases
the calculated SOC constants $\zeta^{avD}(\ell)$,
based on $P_{\ell}^{av}$-functions, give much better energy differences $\triangle_{SO}^{avD}(d)$, $\triangle_{SO}^{avD}(f)$
than $\triangle_{SO}^{KH}(d)$, $\triangle_{SO}^{KH}(f)$, based on the KH functions $P_{l}^{KH}$.
This is directly related to the behavior of the avD and KH radial functions close to the nuclear region,
where the functions $P_{\ell=2}^{av}$ ($P_{\ell=3}^{av}$)
are systematically larger than $P_{l=2}^{KH}$ ($P_{l=3}^{KH}$), Fig.\ \ref{fig3}.
Although the difference between the $d-$ and $f-$ basis functions in the whole region is rather small, Table \ref{tab1},
the larger values of $P_{\ell=2}^{av}$ ($P_{\ell=3}^{av}$) is the decisive factor which finally leads to larger
SO coupling constants and better values for the energy splittings.
%
%

The situation however is changed for the SO interactions of $6p$ states.
The reason for this is a very different radial dependence of the $p_{1/2}$ and $p_{3/2}$ radial components in the nuclear region,
shown in Fig.\ \ref{fig1} and Fig.\ \ref{fig2}, and discussed earlier in Sec\ \ref{sub:basis}.
In the neighborhood of the nucleus, important for SOC, the $p_{1/2}$ component is very large (even singular for the point nucleus),
and as a result, $\zeta_{1/2}$ calculated with the $p_{1/2}$ function is more than six times larger than $\zeta_{3/2}$
calculated with the $p_{3/2}$ component, Table \ref{tab3}.
Even after the averaging between two $p$-components according to Eq.\ (\ref{i1a}), the calculated SOC constant $\zeta^{avD}(p)$
overestimates approximately twice the actual SO-splitting, i.e. for the avD-basis $\triangle_{SO}(p) \approx 2 \times \triangle E(p)$,
where $\triangle E(p) = E(6p_{3/2}) - E(6p_{1/2})$ is the actual splitting.
The use of KH $6p$ radial functions improves the situation but it is also far from being ideal.
Although in the nuclear region the KH-radial $6p$ functions $P_{l=1}^{KH}$ \cite{KH,MPK} is appreciably smaller than the corresponding avD-function $P_{\ell=1}^{av}$,
Fig.\ \ref{fig2}A, its SOC constant $\zeta^{KH}(p)$ remains large.
Inspection of Table \ref{tab3} shows that $\triangle_{SO}^{KH}(p)$ overestimates the actual values $\triangle E(p)$
by 17{\%} for Ac, 19{\%} for Th and 27{\%} for Np.
Earlier the problem of overestimated SO coupling effects was noticed e.g. in Ref.\ \cite{Jones}.
The situation is aggravated by large absolute values of the $6p$-splittings (6.7-9.5~eV), which have a large impact on the band structure calculations.

\begin{figure}
\begin{flushleft}
\resizebox{0.47\textwidth}{!}
{
\includegraphics{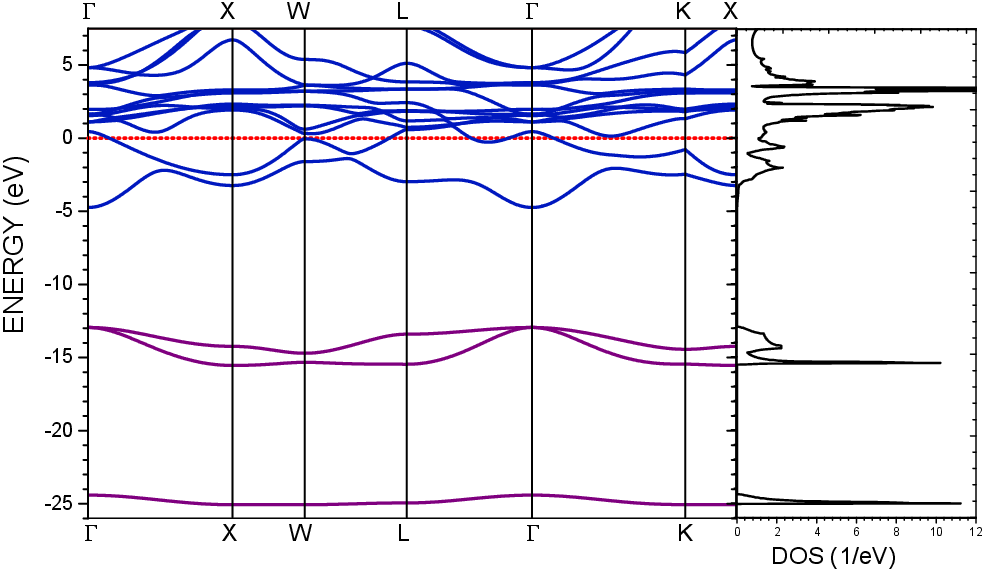}
}
\caption{
Band structure of Th with the full SO coupling (left) and the density of electron states (DOS, right) in the KH radial function basis \cite{KH,MPK}.
Three lowest bands represent the $6p$ semicore states split by the SOC. The middle-point
difference ($10.7$~eV) exceeds the $6p$ splitting in the Th atom (7.8~eV), see text for details.
} \label{fig4}
\end{flushleft}
\end{figure}

From the table~\ref{tab3} we can conclude that the actual energy splittings $\triangle E(p)$ between the $p_{1/2}$ and $p_{3/2}$ states are better approximated
by the SO coupling constant $\zeta(p_{3/2})$ calculated using a single radial component $p_{3/2}$, for which $\triangle_{SO}^*(p)=3 \zeta(p_{3/2})/2$.
For example, for Ac we obtain $\triangle_{SO}^*(p)=6.09$~eV, for Th $\triangle_{SO}^*(p)=7.12$~eV etc.
Although the values $\triangle_{SO}^*(p)$ are slightly smaller than the real energy differences $\triangle E(p)$,
they approximate $\triangle E(p)$ better (i.e. $9-11${\%} vs $17-27${\%}) than the KH radial functions $P_{l=1}^{KH}$.
The overestimation holds also for the solid case. Our data indicate that for example for the canonical (KH) Th variant the SO splitting between the $6p$-subband midpoints
is 10.7~eV (with the minimum and maximum difference being 8.9 and 12.1~eV) whereas in the atomic case it is only 7.8~eV, Fig.~\ref{fig4}.
Note that in the solid or atomic case the effective potential in the nuclear region, which mostly affects the splitting, is basically unchanged.
Therefore, in the following for a more realistic description of the SO splittings for the $6p$ semicore band states within the LAPW method
we suggest to use only $p_{3/2}$ radial component.
Since the actual equations for the SO coupling in LAPW (see e.g. Eq.~(2) in \cite{Nik3}) differ from Eq.\ (\ref{so1}) and Eq.\ (\ref{so1B}),
and require three SO coupling constants $\zeta(p)$, $\dot{\zeta}(p)$, $\ddot{\zeta}(p)$, computed with two functions $P_{l=1}$ and $\dot{P}_{l=1}$,
this implies that for their calculations we use $P_{3/2}$ (the $6p_{3/2}$ large component) and $\dot{P}_{3/2}$, which is the first energy derivative of the $6p_{3/2}$ large component.
%
%

For the other SOC calculations (that is, for the $d$ and $f$ and higher $\ell$ valence states) we
use the standard procedure with the averaged radial components $P_{l}^{av}$ and $\dot{P}_{l}^{av}$, Eq.\ (\ref{i1a}),
because, as discussed above, they give good approximations of the energy splittings in the atomic case (i.e. $\triangle_{SO}(d)$ and $\triangle_{SO}(f)$).
Our results obtained using this SO treatment are presented in the tables~\ref{tab6}--\ref{tab9} below as avD variants.
The KH variants in the Tables correspond to the canonical SO treatment, which as discussed earlier overestimates the $6p$ energy splitting.
Here it is worth mentioning that the effect of overestimation of the $6p$ SO energy splitting is clearly visible in the full treatment of SOC (in the full basis set).
The SOC calculated in the second variation step as a consequence of smaller basis set, gives smaller energies of the $6p$ splitting which can compensate this effect.
Although in this Section we have considered the results with the PBE variant of DFT, the same conclusions can be drawn for other DFT functionals.
%
\begin{table}
\caption{
Calculated SOC constants $\zeta$ (in eV) with the $6d_{3/2}$ and $6d_{5/2}$ radial functions in atoms.
$\zeta(d)$ is the averaged value for two basis sets: avD, Eq.\ (\ref{i1a}),
and KH, \cite{KH,MPK}. $\triangle_{SO}(d)$ (in eV)
is the corresponding SO energy splitting, whereas $\triangle E = E(6d_{5/2}) - E(6d_{3/2})$ (in eV)
is the difference according to the fully relativistic Dirac atomic calculation.
\label{tab4} }
\begin{ruledtabular}
\begin{tabular}{l l  c  c  c  c  c}
     &  basis & $\zeta(d_{3/2})$  & $\zeta(d_{5/2})$ & $\zeta(d)$ & $\triangle_{SO}(d)$ & $\triangle E(d)$  \\
\tableline
 Ac & avD & 0.182 & 0.142 & 0.158 & 0.394 & 0.372 \\
 Ac & KH  &       &       & 0.125 & 0.313 & 0.372 \\
 Th & avD & 0.250 & 0.196 & 0.216 & 0.541 & 0.510 \\
 Th & KH  &       &       & 0.182 & 0.456 & 0.510 \\
 U  & avD & 0.292 & 0.224 & 0.250 & 0.624 & 0.587 \\
 U  & KH  &       &       & 0.212 & 0.531 & 0.587 \\
 Np & avD & 0.254 & 0.188 & 0.214 & 0.534 & 0.502 \\
 Np & KH  &       &       & 0.230 & 0.574 & 0.502 \\
\end{tabular}
\end{ruledtabular}
\end{table}

%
\begin{table}
\caption{
Calculated SOC constants $\zeta$ (in eV) with the $5f_{5/2}$ and $5f_{7/2}$ radial functions in atoms.
$\zeta(f)$ is the averaged value for two basis sets: avD, Eq.\ (\ref{i1a}),
and KH, \cite{KH,MPK}. $\triangle_{SO}(f)$ (in eV)
is the corresponding SO energy splitting, whereas $\triangle E = E(5f_{5/2}) - E(5f_{7/2})$ (in eV)
is the difference according to the fully relativistic Dirac atomic calculation.
\label{tab5} }
\begin{ruledtabular}
\begin{tabular}{l l  c  c  c  c  c}
     & basis &  $\zeta(f_{5/2})$  & $\zeta(f_{7/2})$ & $\zeta(f)$ & $\triangle_{SO}(f)$ & $\triangle E(f)$  \\
\tableline
 Th & avD & 0.198 & 0.180 & 0.188 & 0.658 & 0.650 \\
 Th & KH  &       &       & 0.164 & 0.573 & 0.650 \\
 U  & avD & 0.276 & 0.254 & 0.263 & 0.922 & 0.909 \\
 U  & KH  &       &       & 0.242 & 0.848 & 0.909 \\
 Np & avD & 0.292 & 0.264 & 0.276 & 0.967 & 0.953 \\
 Np & KH  &       &       & 0.283 & 0.992 & 0.953 \\
\end{tabular}
\end{ruledtabular}
\end{table}

%
\begin{table}
\caption{
Calculated SOC constants $\zeta$ (in eV) with the $6p_{1/2}$ and $6p_{3/2}$ radial atomic functions.
$\zeta(p)$ is the averaged value for two basis sets: avD, Eq.\ (\ref{i1a}),
and KH, \cite{KH,MPK}. $\triangle_{SO}$ (in eV)
is the corresponding SO energy splitting, $\triangle^*_{SO} = \triangle_{SO}(p_{3/2})$ (in eV) is the energy splitting for $\zeta(p_{3/2})$,
whereas $\triangle E = E(6p_{3/2}) - E(6p_{1/2})$ (in eV)
is the actual difference according to the fully relativistic Dirac atomic calculation.
\label{tab3} }
\begin{ruledtabular}
\begin{tabular}{l  l  c  c  r  c  r  c}
     & basis &  $\zeta(p_{1/2})$  & $\zeta(p_{3/2})$ & $\zeta(p)$ & $\triangle^*_{SO}$ & $\triangle_{SO}$ & $\triangle E(p)$  \\
\tableline
 Ac    & avD & 26.22 & 4.06 &  8.17 &  6.09 & 12.25 & 6.69 \\
 Ac    & KH  &       &      &  5.22 &       &  7.84 & 6.69 \\
 Th    & avD & 32.16 & 4.75 &  9.78 &  7.12 & 14.66 & 7.80 \\
 Th    & KH  &       &      &  6.18 &       &  9.26 & 7.80 \\
 U     & avD & 42.98 & 5.58 & 12.28 &  8.36 & 18.42 & 9.26 \\
 U     & KH  &       &      &  7.40 &       & 11.10 & 9.26 \\
 Np    & avD & 47.40 & 5.66 & 13.05 &  8.49 & 19.57 & 9.53 \\
 Np    & KH  &       &      &  8.06 &       & 12.08 & 9.53 \\
\end{tabular}
\end{ruledtabular}
\end{table}

In addition to the SO couplings occurring inside the MT-sphere region, we have examined the
SO effect in the interstitial region (IR).
The matrix elements of the SOC there are given by
\begin{eqnarray}
  \langle \phi_p | V^{SO} |  \phi_j \rangle
  = \frac{i}{4 c^2} \sum_{\vec{K}}  F(\vec{K}_j - \vec{K}_p + \vec{K})\,  V_{\vec{K}}         \nonumber \\
   \left[ \vec{K} \times (\vec{k} + \frac{1}{2}(\vec{K}_j + \vec{K}_p)) \right] \vec{\sigma} ,
   \label{so2}
\end{eqnarray}
where $\vec{K}_j$, $\vec{K}_p$ are the corresponding reciprocal lattice vectors,
$V_{\vec{K}}$ is the Fourier component of the potential in IR, $\vec{\sigma}$ are the Pauli matrices,
$F(\vec{K})$ are the standard LAPW integrals of $\exp(i \vec{K} \vec{R}) / v$ in IR.
Our calculations indicate that the effect of the additional SOC in the interstitial region, Eq.\ (\ref{so2}),
is negligible. This is related to the fact that the variations of the total potential in IR are very small
in comparison with the changes near nuclei. Therefore, the interstitial region can be safely considered
as nonrelativistic.

\section{Application to actinides}
\label{sec:appl}

For calculation of the exchange-correlation potential and the exchange-correlation energy contribution within the DFT approach,
we have used (1) the Perdew-Burke-Ernzerhof (PBE) scheme \cite{PBE} of the generalized-gradient approximation (GGA),
(2) PBEsol \cite{PBEsol} which is a variant of PBE,
and (3) the local density approximation (LDA) with the standard ($V_{exc} \sim - \rho^{1/3}$) exchange \cite{Dir} and the PW-correlation \cite{PWcorr}.
For the band structure calculations we have used the Moscow-FLAPW method \cite{Nik2,Nik3} which has been widely used by us before
for the study of chemical bonding elemental solids and compounds.
The technical parameters of numerical calculations are given in Appendix~\ref{A:tech}.
Note, that the calculations with new radial basis functions (avD) have been performed without the additional $p_{1/2}$ local atomic function
as done in the LAPW+$p_{1/2}$ method. As discussed in Sec.\ \ref{sub:basis} that is because the new $6p$ radial function $P^{av}_{\ell=1}$ gains
more weight of the $p_{1/2}$ states and on average can describe correctly the fully occupied $6p$ bands.

The results of our calculations are listed in Table~\ref{tab6} for fcc Ac, Table~\ref{tab7} for fcc Th, Table~\ref{tab8} for cubic ThO$_2$,
and Table~\ref{tab9} for cubic UO$_2$. ThO$_2$ and UO$_2$ are crystallized in the CaF$_2$ structure.
As discussed in the Introduction, first calculations of actinides \cite{review-LMTO,Th-LMTO} and their oxides \cite{ThO2-LMTO,Pet,Ols,Kan}
have been performed with the full potential LMTO method.
Although nowadays the full potential LAPW (FLAPW) study of actinides \cite{Jones,Pen} and dioxides \cite{Terki,Shein} are also available,
{\it the new feature of the present FLAPW calculations is the complete (full) treatment of the SO couplings},
which effectively doubles the dimension of the basis set.
This option makes difference with other FLAPW calculations of actinides \cite{Jones,Pen} where the SO coupling was incorporated
at the second variational level \cite{Kunes,Vona},
which introduces certain uncontrolled approximations \cite{blapw}.
Recently, high precision PBE calculations of actinides and their oxides without SO coupling were reported in Ref.\ \cite{Bosoni}.
They can be compared with our results for the PBE variant with the canonical (the KH basis) calculation without SOC.
Our equilibrium lattice constants for Ac and Th agree with those given in the Supplementary information of \cite{Bosoni}
within 0.2\% and 0.3\%, respectively, while for the dioxides (ThO$_2$, UO$_2$), the difference is 1.2\%. This discrepancy is explained by the use of different technical parameters (MT radii, etc.), local orbitals \cite{Nik2}, and even fitting functions (we use the Murnaghan equation of state).

It is worth mentioning that there are several theoretical studies of UO$_2$ with correlation effects (Hubbard repulsion) \cite{hydr,Shi,Dev,Wang,Vat,UO2-cin,Brun}.
Such an approach however lies beyond the scope of the present work, which focuses on the peculiarities of the inclusion of relativistic effects.

All calculations in the present work are performed for two various basis sets (with the canonical KH radial functions and averaged Dirac [avD] functions, Sec.\ \ref{sub:basis}).
For the avD basis sets we have used the corrected values for $a_l^n$, $b_l^n$, Eq.\ (\ref{cor3a}), Eq.\ (\ref{cor3b}),
and Eq.\ (\ref{cor5}) for $\gamma^l$ for the matrix elements in the spherical ($L=0$) component.
For the KH basis set we have adopted the values given in Eq.\ (16a) and Eq.\ (16b) of Ref.\ \cite{KA}.
In addition, we have carried out calculations with and without the SO coupling.
For the calculation of the SO coupling constant of the $6p$ semicore states in the avD bases we used the $6p_{3/2}$ large component
as described in Sec.\ \ref{sub:soc-p}, and for the KH basis the canonical averaged radial $6p$ component, which overestimates the SO energy splitting,
see more details in Sec.\ \ref{sub:soc-p}.
For the SO coupling constants of other valence states (i.e., $d-$, $f-$, and higher $\ell-$ states), we employed averaged radial functions, which, however, differ
in the avD and KH schemes, Section \ref{sub:soc-p}.

Inspection of Tables \ref{tab6}--\ref{tab9} shows that even within the same DFT functional (LDA, PBE or PBEsol),
various inclusions of relativistic effects lead to very different results for the equilibrium lattice constants and bulk modulii.
In particular, the largest variation of $a$ reaches 0.147~{\AA} for fcc Ac (LDA) although in the case of ThO$_2$ it is only 0.019~{\AA} for LDA
and 0.01~{\AA} for PBE and PBEsol.
The largest difference in $B$ reaches 26.2~GPa for fcc Th (LDA) and 24~GPa for UO$_2$ (LDA),
although, for example, for Ac (PBE) it is only 2.4~GPa.
Inclusion of the SO coupling leads to smaller lattice constants for fcc Ac and Th,
but to larger ones for UO$_2$. As a rule, the SO coupling results in larger bulk moduli, but in some
cases they practically do not change (Ac, PBE and PBEsol; ThO$_2$, all DFT) or even get smaller (Ac, LDA, PBE, PBEsol with avD; or UO$_2$ with all DFT variants).

The opposing trends are also found for the avD and KH basis sets.
In some cases the use of the KH functions leads to smaller lattice constants (fcc Ac and Th), but
in other instances (ThO$_2$ or UO$_2$) it gives larger values of $a$.
The bulk moduli calculated with the KH functions can be larger (Th, all DFT variants; ThO$_2$, LDA), but also smaller than $B$ found with the avD variants
(UO$_2$, all DFT functionals).
The other characteristics of the band structure are also susceptible to different treatment of relativistic effects.
For example, the gap $E_g$ of forbidden states in ThO$_2$ changes by 0.25~eV ($\sim 5${\%}), Table \ref{tab8}.

%
\begin{table}
\caption{
Results of LAPW calculations for fcc structure of elemental actinium (Ac) with the averaged Dirac (avD)
and Koelling-Harmon \cite{KH} (KH) radial basis functions,
with SOC (marked by $*$) and without it.
$a$ is the equilibrium lattice constant (in {\AA}), $B$ is the bulk modulus (in GPa).
Experimental data: $a=5.315$~{\AA} \cite{Ac1}, estimated $B=24.5$~GPa \cite{Bexp}.
\label{tab6} }
\begin{ruledtabular}
\begin{tabular}{c  l  c  c  c }
  DFT  & Basis & SO & $a$ ({\AA}) & $B$, GPa  \\
\tableline
LDA    &  avD    &     & 5.576 &	28.2  \\
LDA    &  avD    & $*$ & 5.540 &	27.6  \\
LDA    &  KH     &     & 5.496 &	31.5  \\
LDA    &  KH     & $*$ & 5.429 &	27.2  \\
PBE    &  avD    &     & 5.756 &	24.6  \\
PBE    &  avD    & $*$ & 5.723 &	24.1  \\
PBE    &  KH     &     & 5.682 &	24.6  \\
PBE    &  KH     & $*$ & 5.611 &	25.9 \\	
PBEsol &  avD    &     & 5.633 &	25.9 \\
PBEsol &  avD    & $*$ & 5.592 &	25.4 \\
PBEsol &  KH     &     & 5.553 &	26.0 \\
PBEsol &  KH     & $*$ & 5.479 &	27.9  \\
\end{tabular}
\end{ruledtabular}
\end{table}

%
\begin{table}
\caption{
Results of LAPW calculations for fcc structure of elemental thorium (Th) with the averaged Dirac (avD)
and Koelling-Harmon \cite{KH} (KH) radial basis functions,
with SOC (marked by $*$) and without it.
$a$ is the equilibrium lattice constant (in {\AA}), $B$ is the bulk modulus (in GPa).
Experimental data: $a=5.0845$~{\AA}, $B=58$~GPa \cite{Th}.
\label{tab7} }
\begin{ruledtabular}
\begin{tabular}{c  l  c  c  c }
  DFT  & Basis & SO & $a$ ({\AA}) & $B$, GPa \\
\tableline
LDA    &  avD    &     & 5.015	& 74.1  \\
LDA    &  avD    & $*$ & 4.996	& 63.7  \\
LDA    &  KH     &     & 4.956  & 82.7   \\
LDA    &  KH     & $*$ & 4.910  & 87.7   \\	
PBE    &  avD    &     & 5.142	& 57.0   \\
PBE    &  avD    & $*$ & 5.119	& 55.6   \\
PBE    &  KH     &     & 5.066	& 57.7   \\
PBE    &  KH     & $*$ & 5.009	& 63.0   \\	
PBEsol &  avD    &     & 5.054  & 58.6   \\
PBEsol &  avD    & $*$ & 5.026	& 60.1   \\
PBEsol &  KH     &     & 4.971	& 61.7   \\
PBEsol &  KH     & $*$ & 4.921	& 69.2   \\

\end{tabular}
\end{ruledtabular}
\end{table}

%
\begin{table}
\caption{
Results of LAPW calculations of uranium dioxide ThO$_2$ (CaF$_2$ structure) with with the averaged Dirac (avD)
and Koelling-Harmon \cite{KH} (KH) radial basis functions,
with SOC (marked by $*$) and without it.
$a$ is the equilibrium lattice constant (in {\AA}), $B$ is the bulk modulus (in GPa).
Experimental data \cite{UO2-exp}: $a=5.6001$~{\AA}, $B=198$~GPa.
\label{tab8} }
\begin{ruledtabular}
\begin{tabular}{c  l  c  c  c  c }
  DFT  & Basis & SO & $a$ ({\AA}) & $B$, GPa &  $E_g$, eV  \\
\tableline
LDA    &  avD    &     & 5.587 &	201.0 &	4.65  \\
LDA    &  avD    & $*$ & 5.592 &	208.6 &	4.54  \\
LDA    &  KH     &     & 5.596 &	232.3 &	4.50  \\
LDA    &  KH     & $*$ & 5.603 &	228.5 &	4.38  \\	
PBE    &  avD    &     & 5.686 &	198.5 &	4.69  \\
PBE    &  avD    & $*$ & 5.687 &	200.0 &	4.59  \\
PBE    &  KH     &     & 5.692 &	198.4 &	4.53  \\
PBE    &  KH     & $*$ & 5.697 &	195.8 &	4.44  \\	
PBEsol &  avD    &     & 5.621 &	215.3 &	4.63  \\
PBEsol &  avD    & $*$ & 5.622 &	216.7 &	4.53  \\
PBEsol &  KH     &     & 5.627 &	215.4 &	4.50  \\
PBEsol &  KH     & $*$ & 5.632 &	211.8 &	4.39  \\
\end{tabular}
\end{ruledtabular}
\end{table}

%
\begin{table}
\caption{
Results of LAPW calculations of uranium dioxide UO$_2$ (CaF$_2$ structure) with with the averaged Dirac (avD)
and Koelling-Harmon \cite{KH} (KH) radial basis functions,
with SOC (marked by $*$) and without it.
$a$ is the equilibrium lattice constant (in {\AA}), $B$ is the bulk modulus (in GPa).
Experimental data \cite{UO2-exp}: $a=5.4731$~{\AA}, $B=207$~GPa.
\label{tab9} }
\begin{ruledtabular}
\begin{tabular}{l  l  c  c  c}
  DFT  & Basis & SO &  $a$ ({\AA}) & $B$, GPa \\
\tableline
LDA    & avD   &     & 5.317	& 279.7	 \\
LDA    & avD   & $*$ & 5.346	& 264.5  \\
LDA    & KH    &     & 5.332	& 275.6	 \\
LDA    & KH    & $*$ & 5.358	& 255.7	 \\	
PBE    & avD   &     & 5.435	& 234.8	 \\
PBE    & avD   & $*$ & 5.468	& 224.3	 \\
PBE    & KH    &     & 5.451	& 231.0	 \\
PBE    & KH    & $*$ & 5.481	& 216.8	 \\	
PBEsol & avD   &     & 5.365	& 259.8	 \\
PBEsol & avD   & $*$ & 5.396	& 245.1	 \\
PBEsol & KH    &     & 5.381	& 255.4	 \\
PBEsol & KH    & $*$ & 5.408	& 237.3	 \\
\end{tabular}
\end{ruledtabular}
\end{table}

It is also worth mentioning that in contrast to Th, our DFT calculations of the fcc structure of Ac appreciably overestimate its lattice constant even for LDA.
This however was also noticed e.g. in Ref.\ \cite{Ac-TB} ($a=5.503$~{\AA} in LDA) and very recently was also confirmed for the PBE LAPW variant without SOC \cite{Bosoni} ($a=5.669$~{\AA}).
Therefore, the effect should be attributed to the peculiarity of
the band structure of this element.
Our calculations indicate that this feature becomes more pronounced with increasing quality of the basis set. For example,
decreasing the non-spherical components of electron density to $L_{max}=6$ in the avD basis leads to a smaller lattice constants: 5.536~{\AA} ($\delta a=-0.037$~{\AA}) in LDA,
5.736~{\AA} ($-0.018$~{\AA}) in PBE, 5.597~{\AA} ($-0.032$~{\AA}) in PBEsol.
Further, decrease of the basis set to only 65 functions results in 5.500~{\AA} (total $\delta a=-0.073$~{\AA} in LDA)
5.608~{\AA} ($-0.146$~{\AA}) in PBE, 5.539~{\AA} ($-0.09$~{\AA}) in PBEsol, compare with Table~\ref{tab6}.
Possibly, some properties of the phonon spectrum and mean square displacements of atoms in solid Ac make the description using poor basis sets
more adequate to the experimental data.
Owing to its scarcity and radioactivity, the experimental bulk modulus of Ac is unknown. To the best of our knowledge
in the literature there is only an estimated (not directly measured) value of 24.5 GPa, listed in Ref.\ \cite{Bexp}.
Theoretical bulk moduli are $B=25.9$~GPa, obtained on the basis of a tight-binding (LDA) analysis \cite{Ac-TB},
and $B=23.9$~GPa in the canonical PBE LAPW variant without SOC \cite{Bosoni}.
All these values are in good correspondence with our data, Table~\ref{tab6}.

\begin{figure}
\resizebox{0.45\textwidth}{!} {
\includegraphics{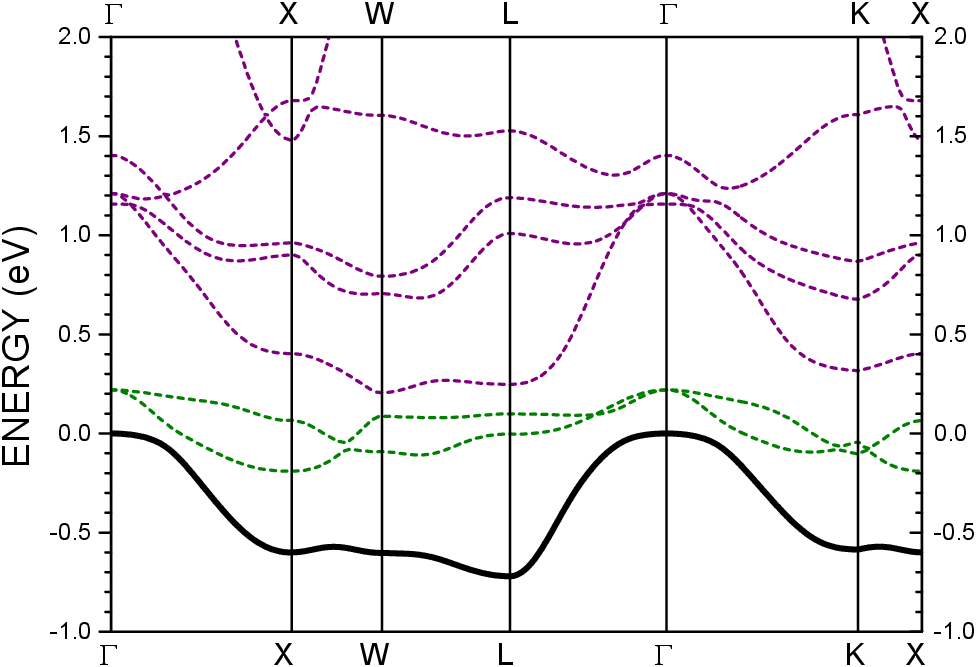}  }

\caption{
The upper panel of the calculated band structure of UO$_2$ with the spin-orbit coupling,
PBE calculation. The highest occupied $5f$ electron band is shown by solid line,
lowest unoccupied $5f$ bands by dashed lines. The vertical gap with $\triangle E$
from $0.2-0.4$~eV is visible.
(The zero energy here corresponds to the position of the first $5f$ band at the $\Gamma$ point, $E_F = -0.16$~eV.)
} \label{fig5}
\end{figure}

Finally, we would like to comment on our calculations of UO$_2$, Table \ref{tab9} and Fig.\ \ref{fig5}.
In particular, it is often stated that in contrast to the experimental observations
the plain band structure analysis predicts the metal character of this compound.
This is not completely correct if the full SOC is taken into account.
As shown in Fig.\ \ref{fig5}, when the full SO coupling is included, at the Fermi energy there is a small gap of $0.2-0.4$~eV between
the highest occupied and the lowest unoccupied $5f$ bands for any chosen $\vec{k}$-vector.
The appearance of the gap in the $5f$ band spectrum can be understood as follows.
In the U atom the $5f_{5/2}$ and $5f_{7/2}$ electron states are split by approximately 1~eV because of the SO interaction.
In the UO$_2$ compound each U atom is surrounded by 8 oxygen neighbors, and the corresponding crystal electric field (CEF) causes
additional energy splittings according to the schemes \cite{Tink},
\begin{subequations}
\begin{eqnarray}
   & &D_{J=5/2} \rightarrow \Gamma_7 + \Gamma_8 , \nonumber \\
   & &D_{J=7/2} \rightarrow \Gamma_6 + \Gamma_7 + \Gamma_8 , \nonumber
\end{eqnarray}
\end{subequations}
In particular, the lowest $J=5/2$ level is split into a doublet ($\Gamma_7$) and a quartet ($\Gamma_8$).
(The $5f$ CEF splittings can be traced at the $\Gamma$ point in Fig.\ \ref{fig5}.)
A small overlap of $5f$ states of U provides the electron band formation,
with the $\Gamma_8$ quartet giving rise to two lowest unoccupied $5f$ bands, but, as follows from Fig.\ \ref{fig5},
a small energy difference between the $5f$ states, originating from the split $\Gamma_7$ and $\Gamma_8$ states, is preserved.
The gap is not clearly observed because the occupied ($\Gamma_7$) and unoccupied ($\Gamma_8$) $5f$ bands slightly overlap at different $\vec{k}$-vectors.
For example, as follows from Fig.\ \ref{fig5} the energy of the $\Gamma_7$ band in the vicinity of the $\Gamma$ point is higher than the energy
of the first $\Gamma_8$ band in the vicinity of the $X$ point.
(In fact, we have performed two different calculations: with the $\Gamma_7$ band, completely occupied, as shown in Fig.\ \ref{fig5}, and with a tiny
occupation of the first $\Gamma_8$ band at the $X$ point [not shown], but the results were virtually the same.)
Therefore, according to our calculations UO$_2$ should be classified as a semimetal.
%
%

However, it is possible that in reality, the inclusion of additional factors (correlations, phonons, etc.) eliminates the slight band overlap, in which case UO$_2$ could become an insulator, consistent with experiments. The situation may be analogous to the calculated band structure of germanium \cite{Asa}, which, according to ref.\ \cite{Asa}, should also be classified as a semimetal. Since germanium is actually a semiconductor, the conclusion about Ge's semimetallic nature is merely an artifact of the calculation.

\section{Conclusions}
\label{sec:con}

Within the LAPW method we have presented a few ways to include the relativistic effects more completely and consistently:
(1) we have used new radial functions $P_{\ell}^{av}$ and $\dot{P}_{\ell}^{av}$ for the Bloch-type basis states. The functions are obtained by finding the large components
$P_{\kappa=\ell}$, $P_{\kappa=-\ell-1}$ of the Dirac solutions independently for the $j = \ell - 1/2$
and $j = \ell + 1/2$ states at the same energy $E_{\ell}$ and then averaging them explicitly by means of Eq.\ (\ref{i1a}).
As discussed in Sec.\ \ref{sub:basis}, the new $6p$ radial functions ($P^{av}_{\ell=1}$ and $\dot{P}^{av}_{\ell=1}$)
gain more weight of the $p_{1/2}$ Dirac solution, and, because of it, on average can describe correctly the fully occupied $6p$ bands even without
the additional $p_{1/2}$ local atomic function as done in the LAPW+$p_{1/2}$ method.
Nevertheless, possibly the new basis functions can also be further enriched with
local atomic-like orbitals \cite{Kunes,Vona,Nik2,Michal,Karsai} as done with the canonical (KH) basis set.

Further,
(2) we have corrected the LAPW expressions for $a_l^n$, $b_l^n$, Eq.\ (\ref{cor3a}), Eq.\ (\ref{cor3b}), and for the matrix elements in the spherical ($L=0$)
component of the total potential, using Eq.\ (\ref{cor5}) for $\gamma^l$.
The canonical expression for $\gamma^l$ given in Eq.\ (16a) and Eq.\ (16b) of Ref.\ \onlinecite{KA} implicitly uses Eq.\ (\ref{cor1}),
which is valid only for the non-relativistic radial solutions $u_l$ and $\dot{u}_l$, Sec.\ \ref{sub:mel};
(3) Since in the full SO treatment the splitting between $6p_{1/2}$ and $6p_{3/2}$ bands is overestimated (see Fig.~\ref{fig4} and the discussion in Sec.\ \ref{sub:soc-p}),
for the calculation of the SOC constants $\zeta(p)$, $\dot{\zeta}(p)$, $\ddot{\zeta}(p)$ for the semicore $6p-$states
we have used the large component of the Dirac solution for the $6p_{3/2}$ states $P_{\kappa=-2}$ and its energy derivative $\dot{P}_{\kappa=-2}$,
which gives better approximation for the actual energy splittings, Sec.\ \ref{sub:soc-p}. We stress that this treatment holds only for the $6p$ states.
For $6d$, $5f$ and high $\ell$ levels, the SOC constants are calculated with the new {\it averaged} radial components $P_{\ell}^{av}$, $\dot{P}_{\ell}^{av}$,
because they describe the SO energy splittings adequately (i.e. here $P_{\ell}^{av}$, $\dot{P}_{\ell}^{av}$
simply substitute the canonical radial functions $P_{\ell}^{KH}$, $\dot{P}_{\ell}^{KH}$ in the LAPW expressions for SOC), Sec.\ \ref{sub:soc-p}.

Our calculations include the SO coupling for valence states in the full basis space,
which differs from other studies where the SO coupling is treated as a second variation step \cite{Kunes,Vona,blapw}.
In addition, in the second variation step method relativistic effects are accounted for by adding relativistic local orbitals (like $6p_{1/2}$) to the LAPW basis set
whereas in our approach they are included in the Bloch-like LAPW basis functions.
The same type of radial functions (i.e. $P_{\ell}^{av}$, $\dot{P}_{\ell}^{av}$) can be used for the construction of local orbital functions
described in Ref.~\cite{Nik2}. Therefore, our approach can be considered as an alternative to the description of relativistic effects within the second variation method.

Based on our study, we have found that the difference in the treatment of relativistic effects
can result in uncertainties up to 0.15~{\AA} for lattice constants and to 26~GPa for bulk moduli
even within the same chosen DFT functional (LDA, PBE or PBEsol), Tables \ref{tab6}--\ref{tab9}.
Unfortunately, as discussed in Sec.\ \ref{sec:appl} it is not possible to conclude on the direction of the changes (i.e. increase or decrease of $a$ or $B$)
when different relativistic treatments are involved: in different materials the trends are opposite.

We have examined the SO coupling in the interstitial region, using Eq.\ (\ref{so2})
for the matrix elements, but the effect appears to be negligible.

Our study of UO$_2$ with the full treatment of the SO coupling clearly demonstrates that it has a small gap ($\sim 0.2-0.4$~eV) at the Fermi energy, Fig.~\ref{fig5},
which persists for all $\vec{k}$-vectors. This is a distinct characteristics of the semimetal, and, therefore,
{\it according to our calculations UO$_2$ is a semimetal.}

We have also found that the DFT calculations of the fcc structure of actinium greatly overestimate its lattice constant for all variants of DFT functionals,
Table~\ref{tab6}.
This effect becomes more pronounced with increasing quality of the basis set, Sec.~\ref{sec:appl}.
The best comparison with the experimental value (with the difference $\triangle a = 0.185$~\AA) is reached in LDA for the basis of 65 basis functions with $L_{max}=6$.

\acknowledgements

This research was supported by a grant of the Russian Science Foundation (Project No 24-12-00053),
and the scientific program of the National Center for Physics and Mathematics, section 6
"Nuclear and radiation physics".

\appendix


\section{Technical parameters of calculations}
\label{A:tech}

The technical parameters of numerical calculations were as follows.
For fcc Ac and fcc Th, in most cases the number of augmented plane waves was 137 and 274 (with SO), with $R_{MT} K_j \sim 10$,
where $\vec{K}_j$ is the maximal wave vector, and $R_{MT}$ is the radius of the MT-sphere.
For CaF$_2$ cubic structures, the basis sets were
307 and 614 (with SO) for ThO$_2$ ($R_{MT} K_j \sim 8.7$), 387 and 774 (with SO) for UO$_2$ ($R_{MT} K_j \sim 9.5$).
For the KH (canonical) basis two sets of $6p$ local functions ($u$ and $\dot u$) as described in Ref.\ \cite{Nik2}.
Therefore, for the KH-calculations the total number of basis functions is 143 and 286 (with SO) for Ac and Th, 313 and 626 (with SO) for  ThO$_2$, and 393 and
786 for UO$_2$.
The following radii of MT-spheres have been used: for Ac 3.46 a.u. for all DFT functionals (i.e. LDA, PBEsol, PBE);
for Th 3.05 a.u. (LDA), 3.12 a.u. (PBEsol and PBE); for ThO$_2$ 2.25 a.u. for both Th and O for all DFT functionals; for
UO$_2$ 2.138 a.u. (LDA), 2.168 a.u. (PBEsol, PBE) for both U and O.
It is worth noticing that within the chosen DFT functional we used the same $R_{MT}$ for all variants of the different radial functions
(i.e. avD, KH with and without SOC).
The linear energy parameter $E_{\ell=1}$ for $p-$states was taken in the midpoint of the semicore $6p$ bands.
The maximal number of $k-$points in the irreducible part (IP) of the Brillouin zone (BZ) for elemental actinides was 1505 ($\sim 70000$ for the whole BZ).
For ThO$_2$ and UO$_2$ we have used a set of 240 $k-$points in IP of BZ ($\sim 11500$ for the whole BZ).
The maximal value of the LAPW plane-wave expansion and the non-spherical density decomposition was $L_{max} = 8$.
We have taken into account the finite size of nuclei and used the tetrahedron method for the linear interpolation of
energy between $k-$points \cite{tet}.
The number of radial points inside the MT sphere region was increased to 4000-4200 for actinides and 900-1000 for oxygen.
The enlarged number of radial points is analogous to the increase of the quality of the basis set, and this is certainly required
for the actinides with 80 core electrons, which is a very large quantity.
The calculations with the KH basis sets have been performed with additional localized atomic $p$-basis functions as described in Ref.\ \cite{Nik2}.
We cannot perform the canonical (KH) calculations without supplying the basis set with local orbitals -- such calculations are plagued with ghost states,
or they crush and do not converge. In the case of new radial basis functions $P_{\ell}^{av}$ (avD) the calculations could proceed regularly until the convergence is reached.
For that reason, all calculations marked in the Tables as avD (with and without SO) have been carried out without local functions.


\end{document}